\documentclass[preprint,3p,times]{elsarticle}

\usepackage{amssymb}
\usepackage{amsmath}
\usepackage{amsthm}
\newtheorem{defi}{Definition}
\usepackage{algorithm}
\usepackage{algorithmic}
\usepackage{multirow}
\usepackage{caption}
\usepackage{subcaption}

\begin{document}

\begin{frontmatter}



\title{RMIO: A Model-Based MARL Framework for Scenarios with Observation Loss in Some Agents}


\author{Shi Zifeng, Liu Meiqin, Zhang Senlin, Zheng Ronghao, Dong Shanling} 

\affiliation{organization={Zhejiang University},
	addressline={No.38 Zhejiang Road, Xihu District}, 
	city={Hangzhou},
	postcode={310058}, 
	state={Zhejiang},
	country={China}}

\begin{abstract}
In recent years, model-based reinforcement learning (MBRL) has emerged as a solution to address sample complexity in multi-agent reinforcement learning (MARL) by modeling agent-environment dynamics to improve sample efficiency. However, most MBRL methods assume complete and continuous observations from each agent during the inference stage, which can be overly idealistic in practical applications. A novel model-based MARL approach called RMIO is introduced to address this limitation, specifically designed for scenarios where observation is lost in some agent. RMIO leverages the world model to reconstruct missing observations, and further reduces reconstruction errors through inter-agent information integration to ensure stable multi-agent decision-making. Secondly, unlike CTCE methods such as MAMBA, RMIO adopts the CTDE paradigm in standard environment, and enabling limited communication only when agents lack observation data, thereby reducing reliance on communication. Additionally, RMIO improves asymptotic performance through strategies such as reward smoothing, a dual-layer experience replay buffer, and an RNN-augmented policy model, surpassing previous work. Our experiments conducted in both the SMAC and MaMuJoCo environments demonstrate that RMIO outperforms current state-of-the-art approaches in terms of asymptotic convergence performance and policy robustness, both in standard mission settings and in scenarios involving observation loss.
\end{abstract}


\begin{highlights}
\item We propose a novel model-based MARL method capable of ensuring stable decision-making even when some agents are completely unable to obtain any observational information.

\item Our approach are the first work to utilize the world model to reconstruct missing observations in multi-agent environment and effectively reduces prediction errors of the world mode by integrating information across agents.

\item Our approach follows the CTDE paradigm in standard (without observation loss) settings and incorporates limited communication through world model to assist decision-making when certain agents cannot access any observations.

\item By adopting reward smoothing and a more reasonable training structure, our method achieves superior experimental results in both standard and scenarios involving observation loss.
\end{highlights}


\end{frontmatter}



\section{Introduction}

Multi-Agent Reinforcement Learning (MARL) provides a powerful and general decision-making framework for multi-agent tasks. By effectively coordinating the interactions between agents, MARL has been extensively applied in tasks that involve both cooperation and competition among agents, such as multi-agent cluster control~\cite{Matignon_Jeanpierre_Mouaddib_2022,Hung_Givigi_2017}, autonomous driving~\cite{You_Lu_Filev_Tsiotras_2019,Shalev-Shwartz_Shammah_Shashua_2016}, and multi-agent games~\cite{rashid2020monotonic,Baker_Kanitscheider_Markov_Wu_Powell_McGrew_Mordatch_2019,ye2020mastering}. In these applications, the partial observability of inputs, the dynamic nature of the environment and policies often necessitate a large number of interaction trajectories to train MARL systems~\cite{Gronauer_Diepold_2022}, which can result in high sampling costs.

Model-based reinforcement learning methods typically simulate real interaction data by constructing an environment interaction dynamics model. This model generates pseudo interaction trajectories, thereby increasing the quantity of sample data and enhancing sample efficiency. This has been confirmed in single-agent reinforcement learning environments~\cite{Hafner_Lillicrap_Ba_Norouzi_2019,Jänner_Fu_Zhang_Levine_2019, Moerland_Broekens_Plaat_Jonker_2023}. Recently, world model techniques based on latent variables have gradually been applied to MARL~\cite{Krupnik_Mordatch_Tamar_2019,Egorov_Shpilman_2022,Wu_Yu_Chen_Hao_Zhuo_2023}. However, the accuracy constraints of the world model in capturing the dynamics of environmental interactions significantly impact the reliability of sample trajectory generation. This hinders the diversified exploration of the real trajectory sample space, making the effective prediction space of the world model narrow and inaccurate~\cite{Wu_Yu_Chen_Hao_Zhuo_2023}.

In the realm of MARL, the predominant methodologies currently define the interaction with the environment as a Decentralized partially observable Markov process (Dec-POMDP)~\cite{Oliehoek_Amato_2016}, under the assumption that each agent receives reliable, albeit limited, observable information at each interaction step. However, in the dynamic and complex real-world scenarios, the observational information obtained by each agent at any moment can be subject to noise~\cite{Sun_Li_Zhao_2012}, time delays~\cite{chen2020delay}, communication limits~\cite{10.5555/2900929.2901013} or even complete loss~\cite{gao2024reinforcement}, often due to communication issues, sensor limitations, or environmental interference. In fact, since the world model is essentially a temporal prediction model, MBRL can utilize it to predict missing or noisy information, making it more suitable for multi-agent tasks under incomplete observation in terms of mechanism. However, much of the current work has primarily focused on using world models to generate data rather than leveraging them to further assist decision-making under incomplete observation conditions.

In this work, we propose a resilient multi-agent optimization framework designed for environments with incomplete observations, called RMIO, stands for \textbf{Robust} decision-making system, \textbf{Model-based} prediction of missing information, and handling of \textbf{Incomplete Observation} challenges. RMIO offers three key contributions:
\paragraph{1. World Model for Completing Missing Observations} RMIO is the first model-based method to utilize the world model to reconstruct missing observations, ensuring stable decision-making even in environments with observation loss. Furthermore, by integrating information across agents, RMIO effectively reduces prediction errors of the world model, further enhancing its ability to handle incomplete observations.
\paragraph{2. CTDE Paradigm} Unlike previous model-based CTCE approaches such as MAMBA~\cite{Egorov_Shpilman_2022}, RMIO adopts the CTDE paradigm in the normal environment, while in the case of some agents are unable to obtain any observations, decision-making is assisted through limited communication.
\paragraph{3. Asymptotic Performance Enhancement} By incorporating strategies such as reward smoothing, a dual-layer experience replay buffer, and adding an RNN network to the policy model, RMIO achieves superior experimental performance in the CTDE paradigm compared to the previous CTCE approach.
     
Experimental results on various tasks in the StarCraftII~\cite{Samvelyan_Rashid_Witt_Farquhar_Nardelli_Rudner_Hung_Torr_Foerster_Whiteson_2019} and MaMuJoCo~\cite{peng2021facmac} benchmark show that RMIO consistently outperforms existing methods in both standard (without observation loss) and observation-loss environment.\textbf{}

\section{Related Workes}
In this section, we discuss the background and related work of RMIO, including the definition of the environment in MARL with probabilistic missing observation information, as well as the technical approaches and recent developments in model-based reinforcement learning (MBRL).
\subsection{MARL}
In most MARL, the problem is typically defined as a Dec-POMDP~\cite{Oliehoek_Amato_2016}. The process is defined by the tuple $\langle N,S,A,P,R,\gamma,\Omega,O\rangle$, where $N$ represents the number of agents, $S$ denotes the global state space of agents,  and $A=\prod{A^i}$ describes the joint action space of agents. $P(\boldsymbol{s}_{t+1}|\boldsymbol{s}_t,  \boldsymbol{a}_t)$ represents the state transition probability function, while $R(\boldsymbol{s}_t, \boldsymbol{a}_t)$ is the reward function that reflects the team reward following the joint action $\boldsymbol{a}_t=\{a^1,...,a^n|a^i\in A^i, i\in\{1,...,n\}\}$ in the state $\boldsymbol{}\in S$. $\gamma \in (0,1]$ is the discount factor which determines the degree of importance given to future rewards. $\Omega(s)$ is the observation space of agents, and $O(s^i)$ is the mapping function from state space $S$ to observation space $\Omega$, which means agent $i$ gets the partial observation $o_t^i$ of $s_t^i$. In the process of Dec-POMDP, the agent $i$ chooses an action $a^i_t$ at time $t$ according to the policy $\pi_i(a^i_t|\tau^i_t)$ , which is conditioned by the action-observation history $\tau^i_t$. Subsequently, the environment will return the team reward $r_t=R(\boldsymbol{s}_t, \boldsymbol{a}_t)$ of the joint action $\boldsymbol{a}_t$ of agents at time $t$. Then the environment will undergoes a transition, with the global state $s^t$ transitioning according to the transition probability function $P(\boldsymbol{s}_{t+1}|\boldsymbol{s}_t,  \boldsymbol{a}_t)$. MARL steps through this process with the goal of maximizing the return of joint policy $\boldsymbol{\pi}=J(\pi^1,...,\pi^n):= \mathbb{E}_{{\pi}}[\sum_{t'=0}^{\infty}\gamma_{t'}r_{t+t'}|\boldsymbol{s}_t, \boldsymbol{a}_t]$. In model-free MARL, $P$ and $R$ are generally not modeled. So the agents need to train their joint policy through frequently interacting with the environment to reach the target.
\subsection{MBRL}
In order to cope with the high sampling cost problem, MBRL employs self-supervised learning to construct an interactive dynamics model, called world model, to estimate the state transition probability distribution $P$ and the reward function $R$. It is proved that expanding sample with the world model can improve sample efficiency~\cite{Jänner_Fu_Zhang_Levine_2019,Feinberg_Wan_Stoica_Jordan_Gonzalez_Levine_2018,Ayoub_Jia_Szepesvári_Wang_Yang_2020}. Recently, considering the intricate nature of dynamic interactions in high-dimensional environments, latent variable world models have been proposed to represent the state transition process in complex scenarios. For instance, the Dreamer series~\cite{Hafner_Lillicrap_Ba_Norouzi_2019, Hafner_Lillicrap_Norouzi_Ba_2020, hafner2023dreamerv3} and related studies leverage the Recurrent State Space Model (RSSM) to represent the state transition process. Conversely, IRIS~\cite{Micheli_Alonso_Fleuret_2022}, Storm~\cite{Zhang_Wang_Sun_Yuan_Huang_2023}, TWM~\cite{Robine} and other related works employ the Transformer~\cite{Vaswani_Shazeer_Parmar_Uszkoreit_Jones_Gomez_Kaiser_Polosukhin_2017} to update their latent state. These methods map the current state information to a latent space, and then perform temporal recursion to estimate the latent state for the next time step. Finally,  the state information at the next time step is reconstructed from the latent space back to the original low-dimensional space. This temporal process enables the simulation of agent-environment interactions and the generation of pseudo trajectories, effectively improving sample efficiency.

While the world model has found extensive application in single-agent tasks, such as Atari games~\cite{Atari}, its utilization in multi-agent environments remains limited. MAMBA~\cite{Egorov_Shpilman_2022}, drawing inspiration from DreamerV2~\cite{Hafner_Lillicrap_Norouzi_Ba_2020}, stands out as a pioneering effort in crafting a world model specifically tailored for multi-agent environments. Based on MAMBA, MAG~\cite{Wu_Yu_Chen_Hao_Zhuo_2023} addresses the issue of local model prediction errors propagating through multi-step rollouts by treating local models as decision-making agents, significantly improving the accuracy of predictions in complex multi-agent environments. Although MAMBA and MAG demonstrate notable improvements in sample efficiency over model-free methods, as CTCE paradigm, their applicability is constrained, and there remains considerable potential for further enhancement in their asymptotic convergence performance. Moreover, these methods primarily use world models for data generation, failing to fully exploit their temporal prediction capabilities. 
\subsection{Incomplete Observations}
Currently, most model-based MARL approaches assume that each agent can accurately obtain local observation information at each time step. However, this assumption overlooks practical limitations in communication and observation systems, such as noisy or missing observations, communication network packet loss, and delays. In real-world interactive environments, agents often cannot acquire complete local observation information in time and may even be entirely deprived of any observation information due to interference. While many studies have addressed these challenges to some extent in model-free MARL methods~\cite{Sun_Li_Zhao_2012, chen2020delay, gao2024reinforcement, Wang_Liu_Li_2020}, there is a notable lack of targeted research in the model-based MARL domain. Therefore, this study focuses on the issue of incomplete observation information to ensure that optimal asymptotic performance is maintained under such conditions. Specifically, we consider scenarios where agents have a probability of receiving no observation information at all, thereby creating extreme observation environments.
\section{Problem Reformulation}

To formalize the environmental challenges under harsh and incomplete information conditions, a detailed definition of this specialized POMDP is provided, as shown in Definition \ref{defi:1}.

\begin{figure}[htbp]
	\centering
	\includegraphics[width=0.6\linewidth]{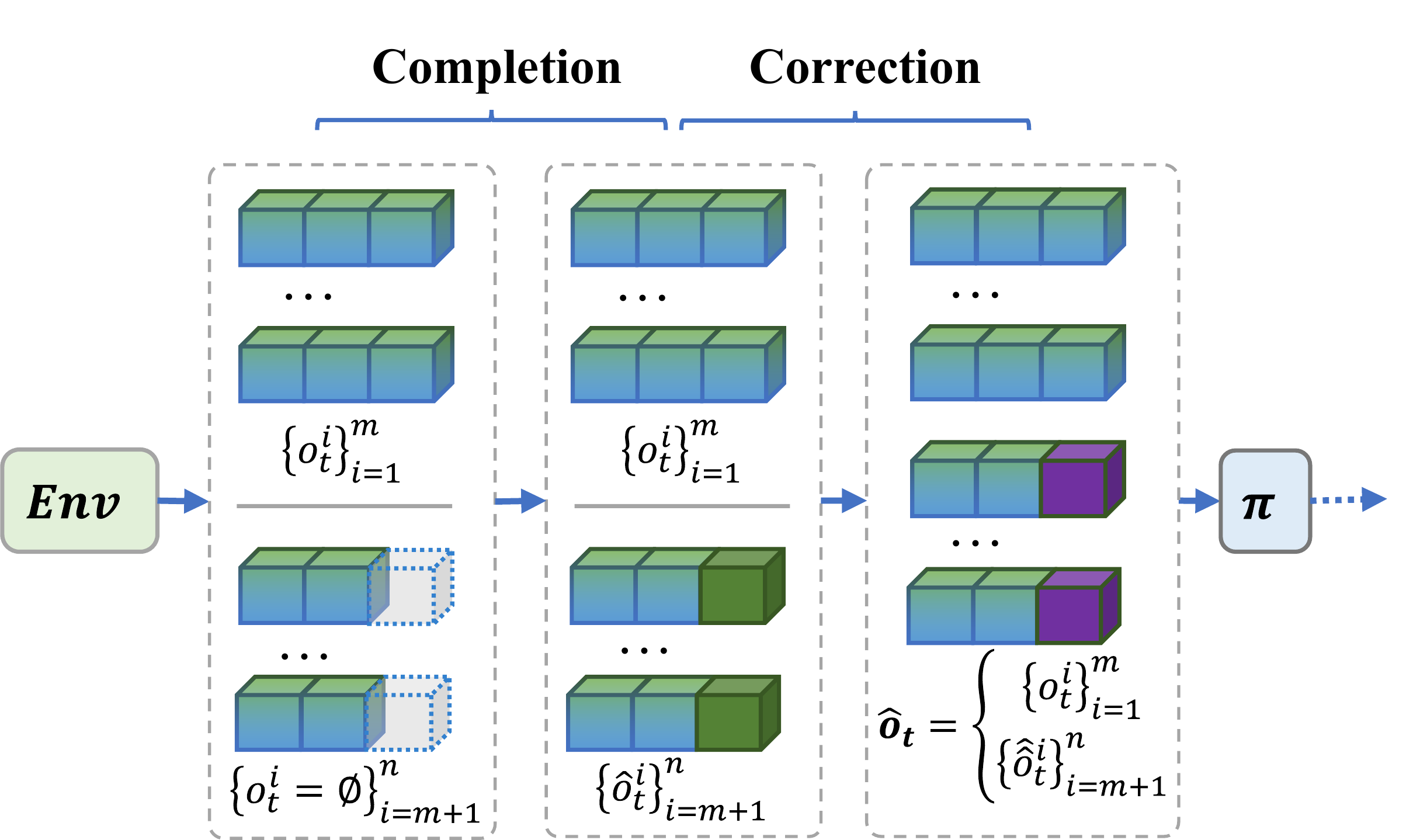}
	\caption{At each step a random subset of agents has a probability of experiencing observation loss. The process of completing and correcting observations by RMIO (only $m$ agents get observations). }
	\label{fig:obsloss}
\end{figure}

\begin{defi}
	\textbf{Observation loss}. At each time step \( t \), agents receive observations \( {o}_t^i \) according to \( O( \boldsymbol{s_t}, {p}_t^i) \), where
	\begin{align}
		O( \boldsymbol{s_t}, {p}_t^i) = 
		\begin{cases}
			o_t^i, &p_t^i,\\
			\varnothing, &1-p_t^i.
		\end{cases}
	\end{align}
	\label{defi:1}
\end{defi}
In order to maintain the stability of the policy under the condition of observations loss, the goal of this work is to ensure the stable performance of both world models and policies in that special environment. As illustrated in Figure~\ref{fig:obsloss}, our method firstly utilizes the world model to fill in missing observation data. Subsequently, it adjusts the $\{\hat{o}_t^i\}^N_{i=m+1}$ based on the reliable observations $\{\hat{o}_t^i\}^m_{i=1}$, aligning it more closely with the actual values,  getting the correction value $\boldsymbol{\hat{o}}_t$. Thereby, we have generalized this special case into a standard POMDP problem.
\section{Methodology}
\label{sec:theory}
In this section, we will introduce the composition of the algorithm and the underlying principles. Next, the improvement in asymptotic performance achieved by RMIO will be discussed. Additionally, the implementation process of RMIO will be deliberated in detail.
\subsection{Architecture}
The architecture of world model, as shown in Formulas (\ref{for:RSSM}) and (\ref{form:pre}), which is improved on the basis of MAMBA, includes RSSM models and predictors.
\begin{align}
&\text{RSSM}\left\{\begin{array}{ll}  
\text{Recurrent model:} \hspace{0.05cm}  &h_t^i = f_{rec}(h_{t-1}^i, e^i_{t-1})\\[2pt]
\text{Posterior model:} \hspace{0.05cm}&z^i_t \sim p_{post}(z^i_t \mid h^i_t, o^i_t)\\[2pt]
\text{Prior model:}\hspace{0.05cm}  &\hat{z}^i_t \sim p_{piror}(\hat{z}^i_t \mid h^i_t)\\[2pt]
\text{Correction block:}\hspace{0.05cm}  &{\hat{o}^i}_t = f_{cor}({\hat{o}}_t^i,o_t^j)\\[2pt]
\text{Communicate block:}\hspace{0.05cm}  &\boldsymbol{e}_t = f_{com}(\boldsymbol{z}_t, \boldsymbol{a}_t)
\end{array}\right. 
\label{for:RSSM}\\
&\text{Predictors}\left\{\begin{array}{ll}
\text{Observation:}        &\hat{o}^i_t \sim p_{obs}(\hat{o}^i_t \mid h^i_t, z^i_t)\\[2pt]
\text{Reward:}      &\hat{r}^i_t \sim p_{rew}(\hat{r}^i_t \mid h^i_t, z^i_t)\\[2pt]
\text{Discount:}    &\hat{\gamma}^i_t \sim p_{dis}(\hat{\gamma}^i_t \mid h^i_t, z^i_t)\\[2pt]
\end{array}\right. 
\label{form:pre}
\end{align}
\subsubsection{RSSM Structure and Reconstruct Predictors}
\paragraph{Recurrent Model} The recurrent model adopts a GRU~\cite{GRU} structure to accurately learn environmental dynamics in partially observable multi-agent environments. It captures historical and current state information through deterministic embeddings \( h_t \) and stochastic embeddings \( z_t \), respectively. During the communication process, the stochastic embeddings \( \boldsymbol{z}_t \) and actions \( \boldsymbol{a}_t \) interact across agents, resulting in \( \boldsymbol{e}_t \), which serves as the input for the recurrent model to update the historical state embeddings.

\paragraph{Posterior Model} The posterior model (representation model) aims to predict \(\boldsymbol{z}_t\) when the observation \(\boldsymbol{o}_t \) is known. This task is potentially easier to learn by minimizing the evidence lower bound~\cite{Kingma_Welling_2013}. In RMIO, the posterior model is additionally designed to update \(\boldsymbol{z}_t\) when there is no observation loss.

\paragraph{Prior Model} In model-based methods, the goal of the prior model (transition model) is to predict $z_t^i$ as accurately as possible without prior information $o_t^i$. It is trained by minimizing the Kullback-Leibler (KL) divergence between \(\hat{z}^i_t\) and \({z}^i_t\) to approximate the posterior model. Thus, the world model can forecast future trajectories without the true observation information and generate samples for training the policy model. Unlike traditional methods, RMIO not only utilizes the prior model to create pseudo trajectories, but also to predict \({z}^i_t\) of agents whose observation is lost. And it then employs the observation predictor to reconstruct the missing observation \(\hat{o}^i_t\) from \({z}^i_t\). The process of completing and predicting missing observations in incomplete environments will be discussed in detail in Section~\ref{sec:algorithm}.

\paragraph{Communication Block} Composed of Transformers~\cite{Vaswani_Shazeer_Parmar_Uszkoreit_Jones_Gomez_Kaiser_Polosukhin_2017}, the communication block  facilitates promoting cross agent learning of global information in world models by integrating the states and actions of different agents, obtaining the encoded embeddings \(\boldsymbol{e}_{t}\). 
\paragraph{Reconstruct predictors} Observation, reward and discount predictors are employed to reconstruct \(\boldsymbol{o}_{t}\), \(r_{t+1}\),  and \(\gamma_{t}\) from \(\boldsymbol{h}_{t}\) and \(\boldsymbol{z}_{t}\). These predictors are trained via supervised loss.

The world model joint loss includes temporal prediction KL divergence loss and predictor reconstruction loss. Minimize the joint loss function through gradient descent to update the world model.
\begin{align}
\label{align:jointlossfunc}
\begin{split}
\mathcal{L}({\theta_{\mathcal{M}}})=\sum_{t=1}^{T}-\ln p\left(\widehat{\boldsymbol{o}}_{t} \mid \boldsymbol{h}_{t}, \mathbf{z}_{t}\right)-\ln p\left(\hat{\boldsymbol{r}}_{t} \mid \boldsymbol{h}_{t}, \mathbf{z}_{t}\right) \\ -\ln p\left(\widehat{\boldsymbol{\gamma}}_{t} \mid \boldsymbol{h}_{t}, \mathbf{z}_{t}\right)+ \beta K L\left[\mathbf{z}_{t}| | \widehat{\mathbf{z}}_{t}\right]
\end{split}
\end{align}
\subsubsection{Correction Block} 
\begin{figure}[htbp]
	\centering
	\includegraphics[width=0.55\linewidth]{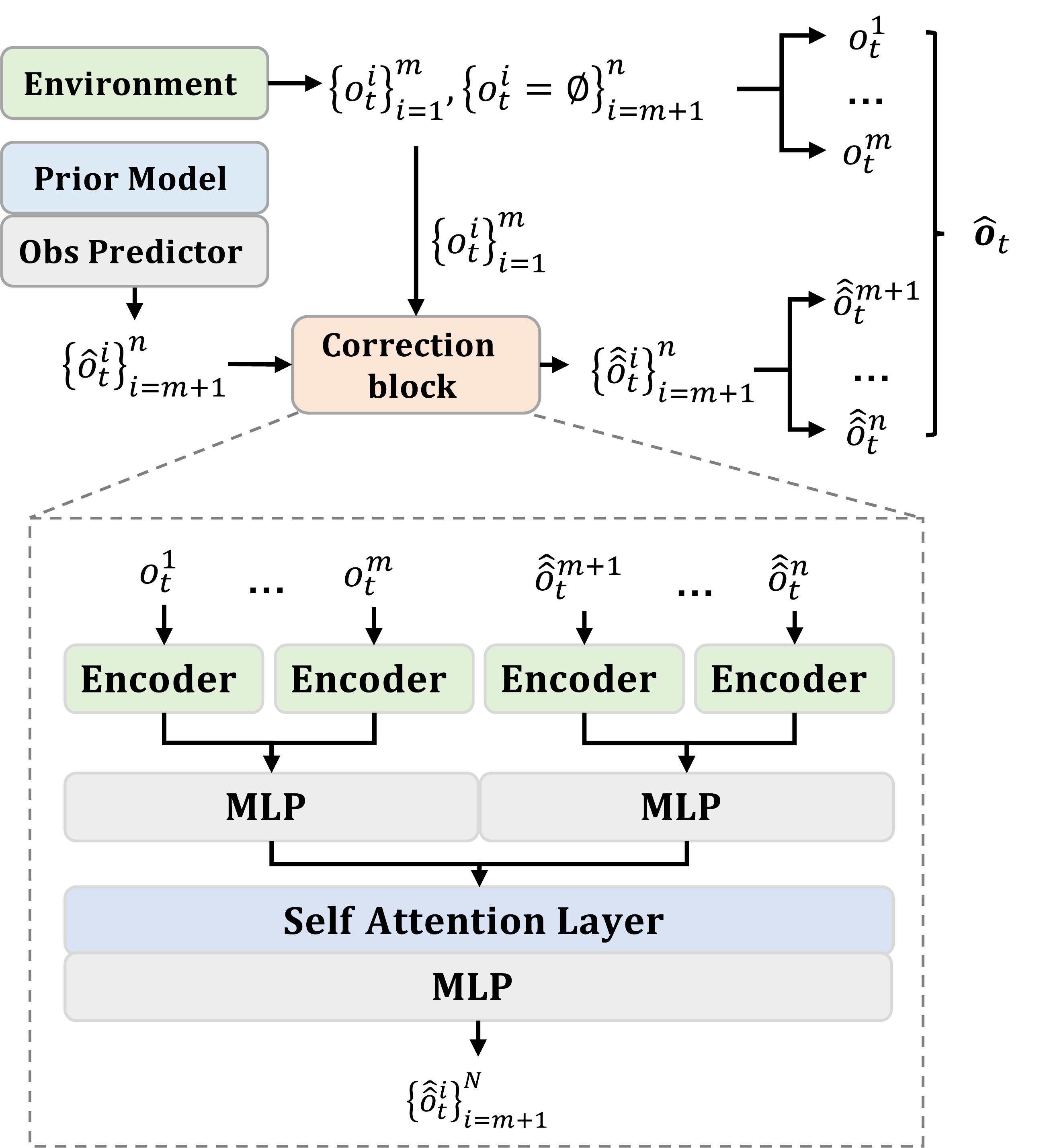}
	\caption{Network composition and inference process of correction block. The scenario in the figure assumes that there are $m$ agents obtaining accurate observations, and $n-m$ agents cannot obtain any observation values.}
	\label{fig:correctionblock}
\end{figure}
When observation loss occurs to several agents (assuming the number is $m$), RMIO leverages the prior model and observation predictor to reconstruct the missing observation and obtain the imputed observation \(\{\hat{o}_t^i\}_{i=m+1}^n\). However, errors are inevitable in this process. To minimize the impact of prior prediction errors, RMIO proposes a correction block to refine the observation estimates obtained from the prior model, as shown in Figure~\ref{fig:correctionblock}. This correction block takes accurate observations \(\{o_t^i\}_{i=1}^m\) and imputed observations \(\{\hat{o}_t^i\}_{i=m+1}^n\) as inputs. It extracts accurate information from \(\{o_t^i\}_{i=1}^m\) to aid in correcting \(\{\hat{o}_t^i\}_{i=m+1}^n\), ultimately generating the corrected values \(\{\hat{\hat{o}}_t^i\}_{i=m+1}^n\). The corrected estimates are then concatenated with the true values to form \(\boldsymbol{\hat{o}}_t\), as illustrated in Formula~\ref{correction}.
\begin{align}
\boldsymbol{\hat{o}}_t &= \{f_{cor}(\{\hat{o}^i_t\}^n_{i=m+1},\{{o}^i_t\}^m_{i=1}),\{{o}^i_t\}^m_{i=1}\}  \notag \\
&= \{\{\hat{\hat{o}}^i_t\}^n_{i=m+1},\{{o}^i_t\}^m_{i=1}\}
\label{correction}
\end{align}

The correction block first uses encoders to encode the original real observation \(\{o_t^i\}_{i=1}^m\) (real values) and reconstructed observation \(\{\hat{o}_t^i\}_{i=m+1}^n\) (estimated values) from different agents, and then extracts feature information through the MLP layer and self attention layer~\cite{Vaswani_Shazeer_Parmar_Uszkoreit_Jones_Gomez_Kaiser_Polosukhin_2017}. Finally, the corrected observation $\{\hat{\hat{o}}^i_t\}^n_{i=m+1}$ is decoded and merged with the original real observation data to obtain the $\boldsymbol{\hat{o}}_t$. During the training process, RMIO mask the complete real observation $\boldsymbol{o}_t$ on the agent dimension, getting \(\{o_t^i\}_{i=1}^m\), and use the world model to reconstruct missing observations, getting \(\{\hat{o}_t^i\}_{i=m+1}^n\). The correction block takes these two items as inputs and calculates  $\boldsymbol{\hat{o}}_t$. To accurately fit \(\boldsymbol{\hat{o}}_t\) to the true observations \(\boldsymbol{{o}}_t\), we use the MSE loss function as shown in Formula (\ref{math:loss}). RMIO trains the correction block by minimizing the objective loss function \(\mathcal{L}_{cor}\).
\begin{align}
\mathcal{L}_{cor}(\theta_{\mathcal{C}}) &=  \frac{1}{n-m} \sum_{i=m+1}^{n} \left( o_t^{i} - \hat{\hat{o}}_t^i \right)^2
\label{math:loss}
\end{align}
\subsubsection{Policy} 
\label{policy}
RMIO adopts the MAPPO method \cite{yu2022surprising} as the policy model \(\boldsymbol{\pi}\), leveraging an Actor-Critic architecture. The Actor (policy) model \(\boldsymbol{\pi}\) is trained by optimizing the following objective function:
\[
\mathcal{L}_{policy}(\theta_{\pi}) = \mathbb{E}_t \left[ \min \left( \rho_t(\boldsymbol{\pi}) \hat{A}_t, \text{clip}(\rho_t(\boldsymbol{\pi}), 1-\epsilon, 1+\epsilon) \hat{A}_t \right) \right],
\]
where \( \rho_t(\boldsymbol{\pi}) \) is the importance sampling ratio of the current and old policies, defined as:
\begin{align}
\rho_t(\theta_{\pi}) = \frac{\boldsymbol{\pi}(\boldsymbol{a}_t | \boldsymbol{o}_t)}{\boldsymbol{\pi}_{old}(\boldsymbol{a}_t | \boldsymbol{o}_t)},
\end{align}
 and \( \hat{A}_t \) is the advantage function, computed using Generalized Advantage Estimation (GAE):
\begin{align}
\label{formula:adv}
\begin{split}
&\hat{A}_t = \delta_t + (\gamma \lambda) \delta_{t+1} + (\gamma \lambda)^2 \delta_{t+2} + \dots \\
&\delta_t = r_t + \gamma V(s_{t+1}) - V(s_t),
\end{split}
\end{align}
where \( \gamma \) is the discount factor, and \( \lambda \) is the GAE parameter.

The Critic (value) model \( V \) is trained by minimizing the following value loss function:
\begin{align}
\mathcal{L}_{value}(\phi_V) = \frac{1}{N} \sum_{t=1}^N \left( V(s_t) - \hat{R}_t \right)^2,
\end{align}
where \( \hat{R}_t \) is the target return for timestep \( t \), computed as:  
\begin{align}
\label{formula:return}
\begin{split}
	\hat{R}_t &= r_t + \gamma r_{t+1} + \gamma^2 r_{t+2} + \dots + \gamma^{T-t} r_T\\
	 &= r_t + \gamma V(s_{t+1}),
\end{split}
\end{align}
where bootstrapping is used to incorporate the value estimate if the trajectory has not terminated at \( t+1 \).

Previous model-based methods, such as MAMBA and MAG, rely on centralized feature representations \(\boldsymbol{h}_t, \boldsymbol{z}_t\) of the world model as the input of policy model during both training and execution. Instead, the policy model \(\boldsymbol{\pi}\) in our approach  directly utilizes distributed, local environment observations \({o}_t^i\) (where \(i \in \{1, \dots, N\}\)) for each agent \(i\) as input in both training and execution process. In addition, RMIO adds GRU units to the policy model to make better use of historical information. This structure of decoupling the world model from the policy model makes our method a CTDE (Centralized Training with Decentralized Execution) approach in the standard settings. The policy is then used to compute the agent’s action distribution \({a}_t^i\sim{\pi}^i({a}_t^i|{o}_t^i)\). Notably, these local observations are reconstructed during training using the centralized world model, while during execution they are directly obtained by the agents interacting with the environment.
\subsection{Asymptotic Performance Improvement}
In order to improve the asymptotic performance, RMIO adopts various strategies as follows.
\subsubsection{Reward Smooth} 
MBRL trains policies by generating hypothetical trajectories with predicted rewards. However, precise reward modeling is difficult to achieve due to the dynamic complexity of the environment interaction. Incorrect rewards will seriously affect the iterative convergence process of the policy $\pi$. Inspired by the human intuition, DreamSmooth~\cite{lee2023dreamsmooth} replaces precise reward prediction with rough estimates of rewards in high complexity and sparse reward environments, including Crafter~\cite{Hafner_2021}, RoboDesk~\cite{kannan2021robodesk} and Shadow Hand~\cite{ShadowHand}. Since MARL environments have similar characteristics, RMIO also adopts temporal smooth to the team reward in each episode and ensure that the total rewards remain consistent:	
\begin{align}
\label{rewardSmooth}
\hat{r}_t\leftarrow f(r_{t-H:t+H}) = \sum_{i=-H}^Hf_i\cdot r_{clip(t+i,0,T)} 
\end{align}
where $T$ and $H$ denote the horizons of episode and smoothing, and the smoothing function $f_i$ satisfies $ \sum_{i=-H}^{H}f_i=1$. This method aims to smooth the reward data in each episode, and then use the processed reward data to train the reward model, so that the reward model fits onto the smoothed reward distribution. In our experiments, Gaussian smoothing is chosen to process the reward function in the time series, as shown in Formula~\ref{formula:rewardsmooth}. The use of smoothed rewards in MARL also does not change the optimality of the strategy, and the proof process can be referred to the Appendix.
\begin{align}
\label{formula:rewardsmooth}
f_i = \frac{\exp\left(-\frac{i^2}{2\sigma^2}\right)}{\sum_{i=-H}^{H} \exp\left(-\frac{i^2}{2\sigma^2}\right)}
\end{align}
\subsubsection{Double Experience Replay Buffer}
Due to factors such as overfitting and deviations between the distributions of samples from different batches, the world model may exhibit abnormal iteration periods where the predicted distribution significantly deviates from the true trajectory distribution. These abnormal pseudo trajectories generated during this period, especially the reward samples~\cite{Wang_Liu_Li_2020}, can cause the policy network to optimize in conflicting directions, thereby disrupting the normal convergence optimization process.
\begin{figure}[htbp]
	\centering
	\includegraphics[width=0.6\linewidth]{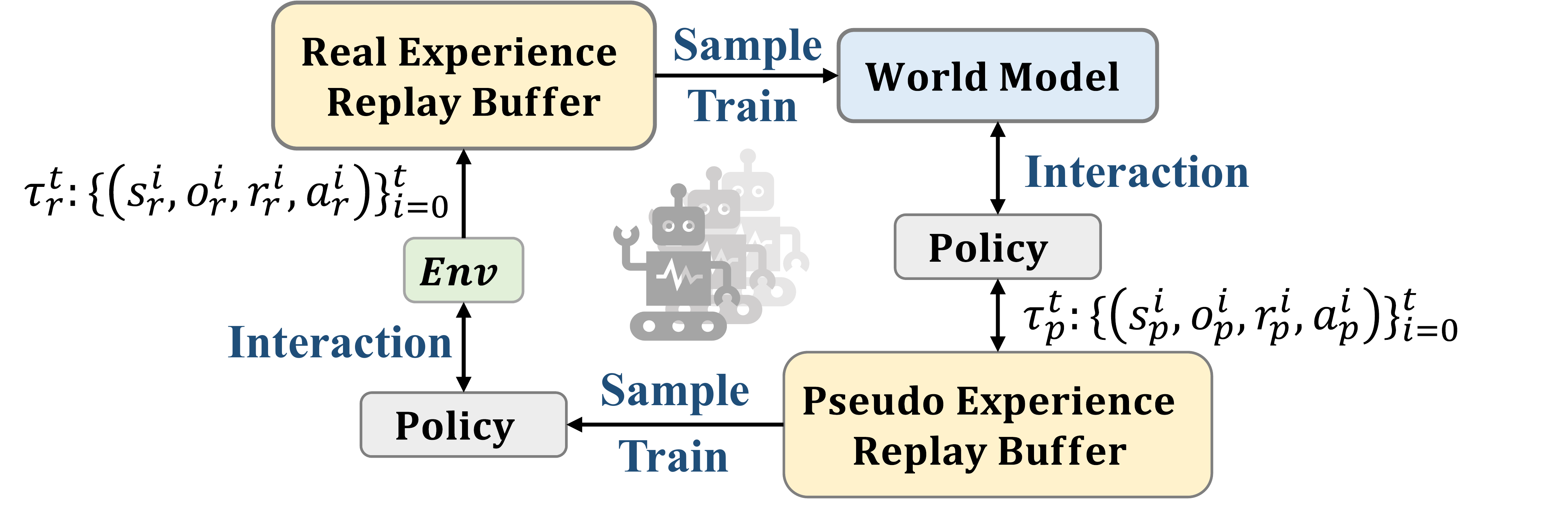}
	\caption{Double Experience Replay Buffer structure.}
	\label{fig:drb}
\end{figure}

In such unstable scenarios, a dual experience replay buffer structure is designed in RMIO to mitigate this issue. As shown in Figure~\ref{fig:drb}, an additional pseudo trajectory experience replay buffer is introduced alongside the original real experience replay buffer for true trajectories. This design reduces the correlation between samples and smooths out changes in the target distribution during training. In addition, compared to training directly on samples generated from a single trajectory fragment, the additional replay buffer contains samples generated from fragments across several different trajectories. This diversity helps prevent the policy from overfitting to the data generated by the single trajectory and enhances its generalization ability. The proof of the effectiveness of the double replay buffer structure can be found in the Appendix.
\subsection{Overall Algorithm Process}
In this section, we provide a detailed explanation of how RMIO manages environments with observed information losses. 
\subsubsection{Training Process of RMIO}
\begin{algorithm}[H]
	\caption{The training process of RMIO}
	\label{alg:train}
	\begin{algorithmic}[1] 
		\STATE Initialize joint policy $\boldsymbol{\pi}$, world model $\mathcal{M}$, correction block $\mathcal{C}$, real trajectory replay buffer $\mathcal{B}_r$ and pseudo trajectory replay buffer $\mathcal{B}_p$.
		\FOR{$N$ episodes}
		\STATE Collect an episode of real-environment trajectory and add it to $\mathcal{B}_r$;
		\FOR{$E_{wm}$ epoches}
		\STATE  Initialize $\boldsymbol{z}_t$ and $\boldsymbol{h}_t$.
		\STATE  Sample $\tau_r= <\boldsymbol{o}_t,\boldsymbol{a}_t,r_t,\boldsymbol{\gamma}_t,\boldsymbol{o}_{t+1}>$ from $ \mathcal{B}_r$.
		\STATE Use $\mathcal{M}$ for one-step temporal prediction and reconstruction on $\tau_r$;
		\STATE Calculate the joint one-step loss :$	\mathcal{L}_\mathcal{M}(\theta_{\mathcal{M}})=\mathcal{L}_{rec} + \beta\mathcal{L}_\mathcal{KL}\notag$;
		\STATE Minimize $\mathcal{L}_\mathcal{M}(\theta_{\mathcal{M}})$ by gradient descent and update $\mathcal{M}$;
		\ENDFOR
		\FOR{$E_\pi$ epoches}
		\STATE Initialize $\boldsymbol{z}_t$ and $\boldsymbol{h}_t$.
		\STATE Sample $\boldsymbol{o_t}$ from $\mathcal{B}_r$ as the initial data. 
		\FOR{$k$ rollout steps}
		\STATE Agents take action $\boldsymbol{a}_t$ according to $\boldsymbol{\pi}(\boldsymbol{a}_t|\boldsymbol{o}_t)$ and communicate to get $\boldsymbol{e}_t = f_{com}(\boldsymbol{z}_t, \boldsymbol{a}_t)$.
		\STATE $\mathcal{M}$ predicts $\{\boldsymbol{o}_{t+1},{r}_{t+1},\boldsymbol{\gamma}_{t+1}\}$, and store them to $\mathcal{B}_p$;
		\STATE Let $\boldsymbol{o}_{t+1}=\boldsymbol{o}_t, t=t+1$;
		\ENDFOR
		\FOR{$E_{sample}$ epoches}
		\STATE Sample $\tau_p=<\boldsymbol{o}_t,\boldsymbol{a}_t,r_t,\boldsymbol{\gamma}_t>$ from $\mathcal{B}_p$; 
		\STATE Compute $\boldsymbol{A}_t$ and returns on $\tau_p$ and compute $\mathcal{L}_{\boldsymbol{\pi}}(\theta_{\pi}),\mathcal{L}_{value}(\phi_V)$;
		\STATE Minimize $\mathcal{L}_{\boldsymbol{\pi}}$ by gradient descent and soft update $\boldsymbol{\pi}$;
		\ENDFOR
		
		\ENDFOR
		\FOR{$E_c$ epoches}
		\STATE Sample continuous $\tau_c=<\boldsymbol{o_t},\boldsymbol{a}_t>$ from $\mathcal{B}_r$. Initialize  $\boldsymbol{z}_t$ and  $\boldsymbol{h}_t$.
		\FOR{$l$ rollout steps}
		\STATE Mask partial agents' observation and estimate prior states $\hat{\boldsymbol{z}}_t=p_{prior}(\boldsymbol{h}_t)$; 
		\STATE
		Reconstruct  $\{\hat{o}_t^i\}_{i=m+1}^n=p_{obs}(\{h_t^i,\hat{z}_t^i\}_{i=m+1}^n$);
		\STATE Use $\mathcal{C}$ to correct $\boldsymbol{\hat{o}}_t=f_{cor}(\{o_t^i\}^m_{i=1},\{\hat{o}_t^i\}^n_{i=m+1})$;
		\STATE Calculate correction loss $\mathcal{L}_{cor}(\theta_{\mathcal{C}})=\text{MSE}(\boldsymbol{{\hat{o}}_t},\boldsymbol{o}_t)$;
		\STATE Predict posterior state $\boldsymbol{z}_t=p_{post}(\boldsymbol{h}_t, \boldsymbol{o}_t)$ and update $\boldsymbol{h}_{t+1}=f_{rec}(f_{com}(\boldsymbol{z}_t,\boldsymbol{a}_t),\boldsymbol{h}_t)$; 
		\ENDFOR
		\STATE Minimize $\mathcal{L}_{cor}(\theta_{\mathcal{C}})$ by gradient descent and update $\mathcal{C}$.
		\ENDFOR
		\ENDFOR		
	\end{algorithmic}
\end{algorithm}
\label{sec:algorithm}

\begin{figure*}[h]
	\centering
	\includegraphics[width=0.9\linewidth]{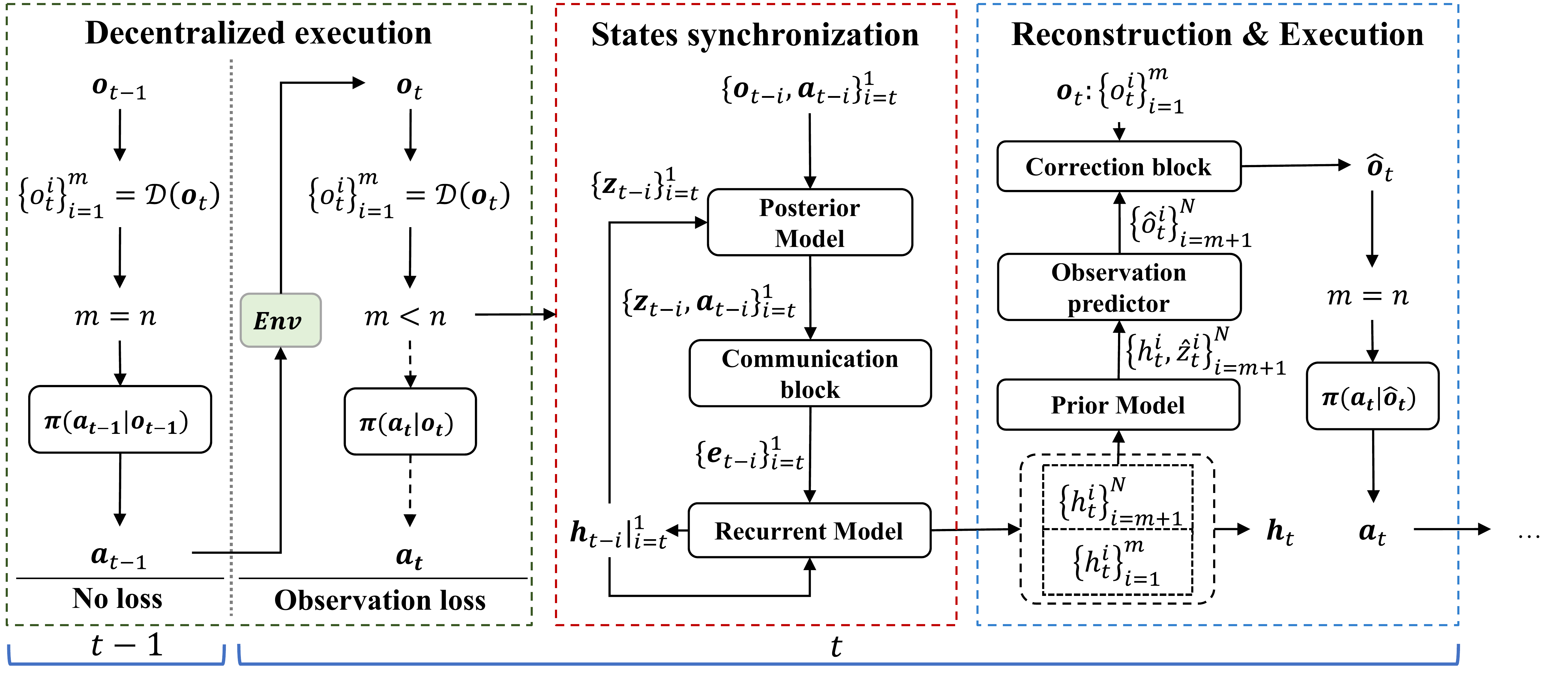}
	\caption{The whole reasoning process of RMIO facing observation loss(only $m$ agents get accurate $\{o_t^i\}_{i=1}^m$) at time step $t$. RMIO first communicate to synchronize the historical $\{\boldsymbol{o}_{t-i},\boldsymbol{a}_{t-i}\}_{i=t}^1$ among agents, getting historical status information $\boldsymbol{h}_t$. Based on $\boldsymbol{h}_t$, the prior model and observation predictor can reconstruct the missing $\{o_t^i\}_{i=n-m+1}^n$, getting $\hat{o}_t^i$. At last, the correction block use the partial accurate $\{{o}^i_t\}_{i=1}^m$ to correct the estimated observation$\{\hat{o}_t^i\}_{i=m+1}^n\}$, getting $\boldsymbol{\hat{o}}_t=\{\{{o}^i_t\}_{i=1}^m,\{\hat{\hat{o}}^i_t\}_{i=m+1}^n\}$. Thus, the agents can take action based on the estimated $\boldsymbol{\hat{o}}_t$. }
	\label{fig:black}
\end{figure*}

As shown in Algorithm~\ref{alg:train}, the training process can be divided into three parts: training world model $\mathcal{M}$ (as shown in lines 3-10 in Algorithm~\ref{alg:train}), training policy $\boldsymbol{\pi}$ (as shown in lines 11-23 in Algorithm~\ref{alg:train}) and training correction block $\mathcal{C}$ (as shown in lines 25-35 in Algorithm~\ref{alg:train}). In the training process of $\mathcal{M}$, RMIO samples from the real sample experience replay buffer $\mathcal{B}_r$ and updates $\mathcal{M}$ by minimizing the joint loss function of single step temporal prediction and state reconstruction through gradient descent, as shown in Formula~\ref{align:jointlossfunc}; During the training process of the policy model $\boldsymbol{\pi}$, RMIO first uses $\mathcal{M}$ to generate pseudo sample trajectories and puts them into the experience replay buffer $\mathcal{B}_p$. Then RMIO samples trajectories in $\mathcal{B}_p$, calculate the policy advantage function and cumulative return on the trajectories according to Formula~\ref{formula:adv} and~\ref{formula:return}, and soft update $\boldsymbol{\pi}$. Since the exploration space of the world model $\mathcal{M}$ is determined by the action decisions generated by the policy model $\boldsymbol{\pi}$, and the quality of the training samples for $\boldsymbol{\pi}$ depends on the prediction accuracy of $\mathcal{M}$, the training processes of the $\mathcal{M}$ and $\boldsymbol{\pi}$ process complement each other. To ensure the stable convergence of both models, model-based methods usually employ an alternating training approach to jointly train the world model and the policy model. The training process of $\mathcal{C}$ only needs to mask the real observations of some agents to simulate the situation of observation loss, as shown in the line 28 in Algorithm~\ref{alg:train}. Then the MSE loss of the correction process is calculated according to Formula~\ref{math:loss} and the weights of $\mathcal{C}$ are updated, as shown in line 34 in Algorithm~\ref{alg:train}.
\subsubsection{Reasoning Process of RMIO}
\begin{algorithm}[h]
	\caption{The reasoning process of RMIO}
	\label{alg:reason}
	\begin{algorithmic}[1] 
		\STATE Load the weights $\theta_w$ of $\mathcal{M}$, weights $\theta_\pi$ of $\boldsymbol{\pi}$; 
		\FOR{$t=t_0;t<T_{done};t=t+1$}
		\STATE Agents get partial observation: $\{o_t^i\}^m_{i=1} = \mathcal{D}(\boldsymbol{o}_t)$;
		\IF{$m<n$}
		\STATE Agents communicate to synchronize $\{\boldsymbol{o}_{h},\boldsymbol{a}_{h}\}_{h=t-l}^{t}$;
		\STATE Initialize $\boldsymbol{h}_{t-l}$ and $\boldsymbol{z}_{t-l}$; 
		\FOR{$h=t-l;h<t;h=h+1$}
		\STATE $\boldsymbol{h}_{h+1}=f_{rec}(f_{com}(\boldsymbol{z}_{h},\boldsymbol{a}_{h}),\boldsymbol{h}_{h})$;
		\ENDFOR

		\STATE Predict stochastic state $\hat{\boldsymbol{{z}}}_t$ : $\hat{\boldsymbol{z}}_t=p_{prior}(\boldsymbol{h}_t)$;
		\STATE Reconstruct $\{{\hat{o}}_t^i\}^n_{i=m+1}=p_{obs}(\{h_t^i,\hat{z}_t^i\}_{i=m+1}^n$);
		\STATE Correct $\boldsymbol{\hat{o}}_t=f_{cor}(\{o_t^i\}^m_{i=1},\{\hat{o}_t^i\}^n_{i=m+1})$;
		\STATE Update stochastic state $\boldsymbol{z}_t=p_{post}(\hat{\boldsymbol{o}}_t,\boldsymbol{h}_t)$;
		\STATE Agents take actions: $a_t^i\sim\pi^i(a_t^i|\hat{o}^i_t)$;
		\ELSE
		\STATE Agents take actions: $a_t^i\sim\pi^i(a_t^i|o^i_t)$;
		\ENDIF
		\ENDFOR
	\end{algorithmic}
\end{algorithm}
After finishing the centralized training process, RMIO can adapt to the scenarios involving observation loss, which is shown in Algorithm~\ref{alg:reason} and Figure~\ref{fig:black}. At each time step, all agents are expected to receive feedback from the environment and utilize the discriminator $\mathcal{D}$ to ascertain if any agents encounter observation loss (as shown in line 3 of Algorithm~\ref{alg:reason}). It is presumed that at time step $t$, $n-m$ agents encounter observation loss, implying that only $m(m<n)$observations$\{o_t^i\}_{i=m+1}^n$. However, in the normal step-by-step process, the joint policy model requires complete observation $\boldsymbol{o}_t$ as inputs to make action decisions. So it is necessary to estimate the missing observations$\{o_t^i\}_{i=m+1}^n$. As shown in lines 5-9 of Algorithm~\ref{alg:reason}, RMIO first uses single-step communication among agents to synchronize  $\boldsymbol{o}_i,\boldsymbol{a}_i$ from the \( l \) steps prior to time \( t \) and applies temporal recursion through the posterior model to obtain deterministic historical state \( \boldsymbol{h}_t \). Based on this, the prior model is used to estimate the stochastic latent state variable \(\hat{\boldsymbol{z}}_t\), and the observation predictor reconstructs the missing observations \(\{\hat{o}_t^i\}_{i=m+1}^n\). Finally, the correction module utilizes the observations of other agents to refine the estimated values of the missing observations, thus generating a joint observation \(\hat{\boldsymbol{o}}_t\) that includes all agents' observation information. For time steps without observation loss, RMIO directly uses the posterior model to obtain the stochastic latent state variable \(\boldsymbol{z}_t\). Unlike traditional model-based MARL, RMIO can maintain a stable action policy regardless of whether agents receive complete observation information.

\begin{figure}[H]
	\centering
	\includegraphics[width=\linewidth]{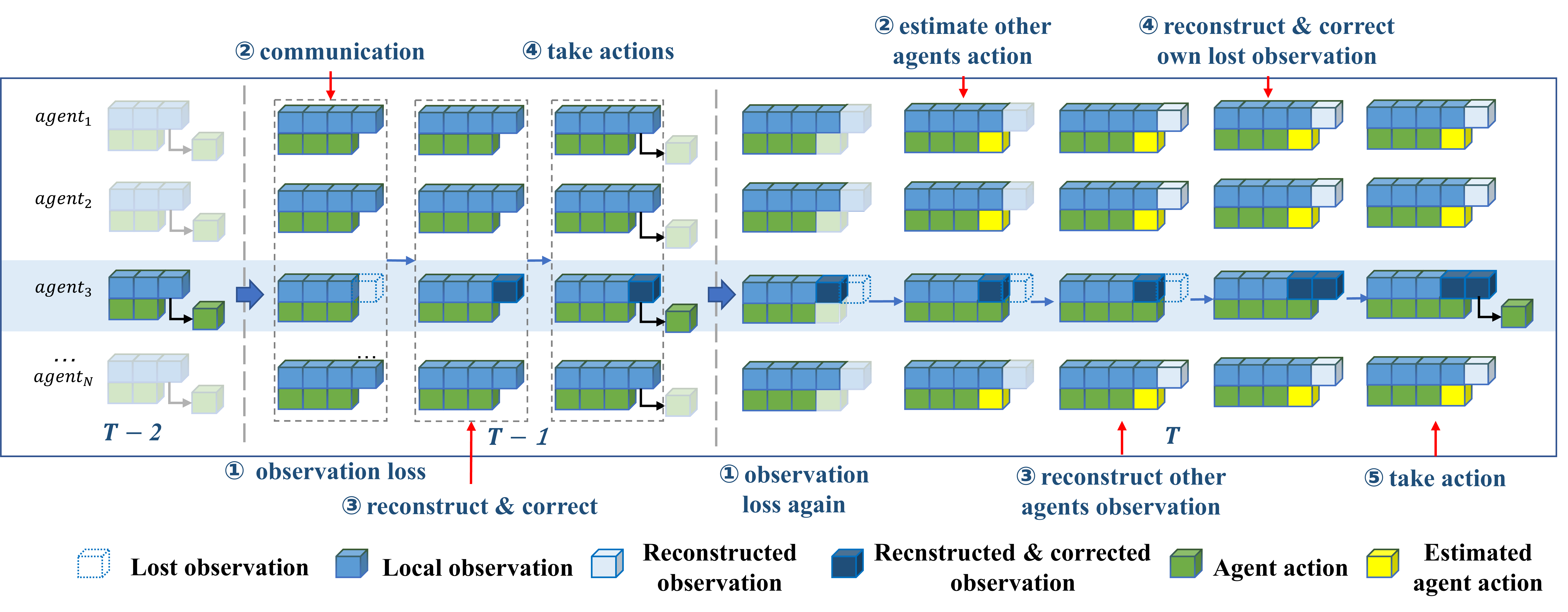}
	\caption{Taking agent 3 as an example, the light colored parts in the figure represent the status information of other agents that agent 3 cannot access. At time \(T-2\), each agent executes in a distributed manner, and agent 3 can only access local state information. At time \(T-1\), agent 3 experiences observation loss. At this time, the missing observation information is supplemented and corrected through communication with the world model to support agent 3's decision-making. Through communication, agent 3 can access some state information of other agents. At time \(T\), agent 3 experiences observation loss again, and it can be considered that the two observation losses are in a "close" proximity state. Agent 3 does not need to communicate at this time, but directly uses the state information of other agents obtained after \(T-1\) communication to make prior estimates of the observed state at time \(T\) and the actions of other agents, thereby completing and correcting the missing observed state at time \(T\). This process does not require further communication and is completed locally by agent 3.}
	\label{fig:comcost}
\end{figure}

As outlined in Section~\ref{policy}, following the CTDE paradigm, RMIO eliminates the need for inter-agent communication in standard environments. Communication is only required in cases of observation loss, where agents perform a single communication step to synchronize historical states, unlike MAMBA and MAG, which rely on continuous communication at every step. Consequently, the communication overhead in RMIO is solely determined by the frequency of observation loss. However, when two instances of observation loss occur in "close" succession, the second instance does not require re-synchronization of historical state information. This is because our agents are designed to share the same set of network parameters, allowing local policy models to estimate the policies of other agents. For agent \(i\), given the observation information of other agents at time \(T-1\), and knowing that the action probability distribution of all agents depends solely on their observed states, it is possible to directly estimate the action probability distribution of other agents as \(a_{T-1}^i \sim \pi(a_{T-1}^i|o_{T-1}^i)\). With the observation \(\boldsymbol{o}_{T-1}\) and action \(\boldsymbol{a}_{T-1}\) at time \(T-1\), the world model can be used to perform a prior estimation of \(o_T\) at time \(T\), and ultimately refine \(o_T^i\). In that case, as shown in Figure~\ref{fig:comcost}, the missing local observations can be completed locally without additional communication by combining the prior predictions from the world model with policy estimates from the policy model. This approach further reduces communication overhead, ensuring that communication frequency is smaller than the frequency of observation loss. In experiment, the standards of "close"  are dynamically adjusted according to the complexity of the experimental environment.

\section{Experiments}

In this section, we will introduce RMIO's empirical research on the challenging StarCratII benchmark (SMAC). In the first part, several baselines (without observation loss) will be compared with RMIO in a normal environment. Subsequently, a quantitative comparative experiments will be conducted to compare the performance retention levels of RMIO and other baselines under different observation loss conditions.
\subsection{Environments} 
The Starcraft Multi Agent Challenge     (SMAC)~\cite{Samvelyan_Rashid_Witt_Farquhar_Nardelli_Rudner_Hung_Torr_Foerster_Whiteson_2019} is a multi-agent discrete and collaborative control benchmark based on StarcraftII. Each task contains a scenario where there are two opposing teams, one controlled by the game robot and the other controlled by our algorithm. The goal is to defeat all the enemy agents. Our method and other baselines are tested on 8 maps of SMAC from \textit{easy} to \textit{super hard}, including \textit{2s\_vs\_1sc, 3s\_vs\_3z, 2s3z, 3s\_vs\_4z, 3s\_vs\_5z, 1c3s5z, 8m, corridor}.

The Multi-Agent MuJoCo (MaMuJoCo)~\cite{peng2021facmac} is a multi-agent continuous and collaborative control benchmark based on the MuJoCo physics simulator. Each task involves a collaborative scenario where multiple agents, represented as different parts of a robot, must work together to complete a specific objective, such as locomotion or manipulation. The goal is to maximize the collective reward by achieving efficient and coordinated control among the agents. Our method and other baselines are evaluated on several tasks from the MAMuJoCo benchmark at different levels of difficulty, including \textit{HalfCheetah} (2 agent), \textit{HalfCheetah} (6 agent), \textit{Swimmer} (2 agent), \textit{Swimmer} (10 agent).
\subsection{Reward Modeling Experiment}
\begin{figure}[ht]
	\centering
	\begin{subfigure}[t]{0.45\textwidth}
		\centering
		\includegraphics[width=0.9\linewidth]{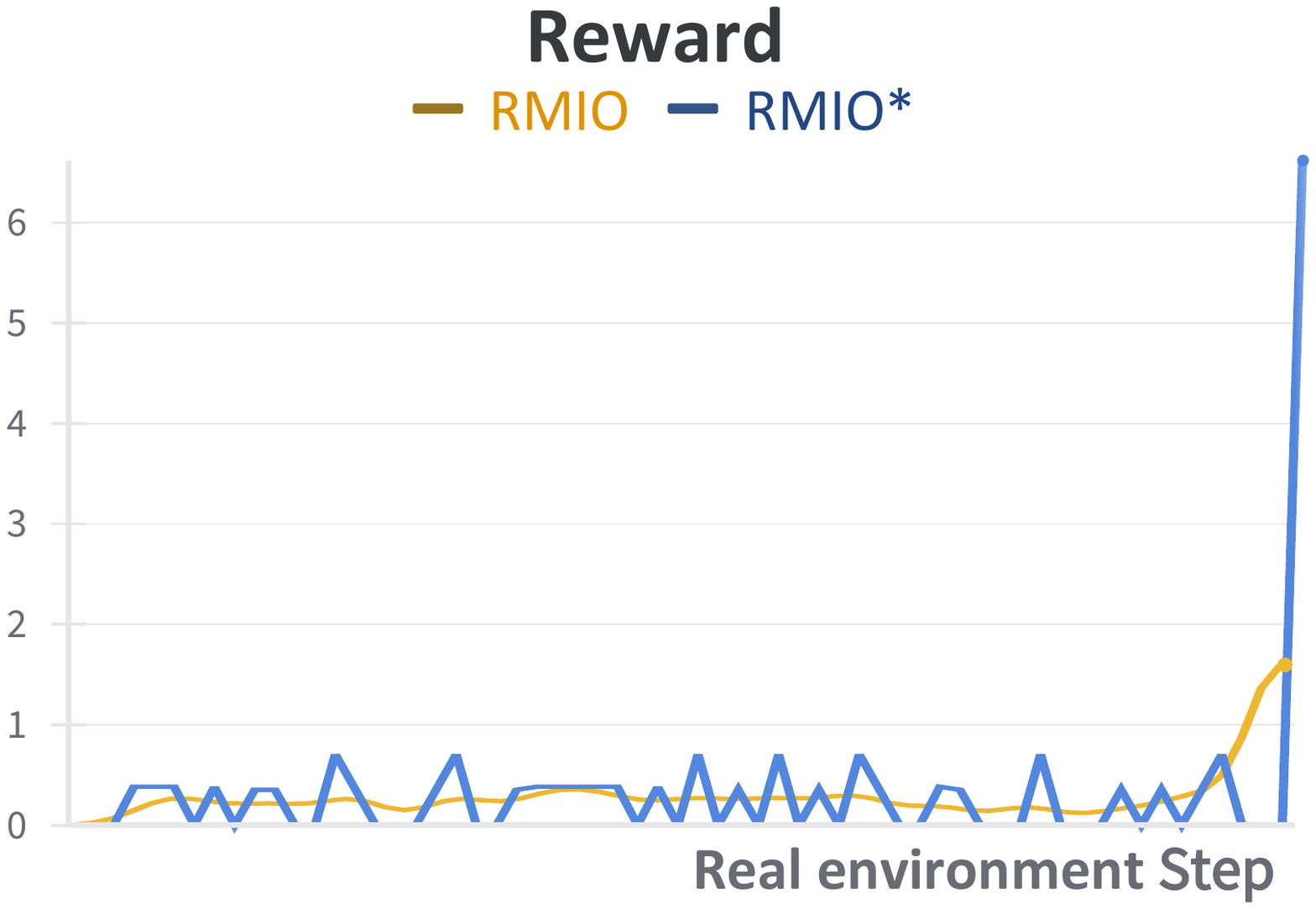}
		\caption{The values of the raw rewards (RMIO*) and smoothed rewards (RMIO) at each time step within a single complete trajectory on the \textit{3s\_vs\_3z} map.}
		\label{fig:Rewardvalue}
	\end{subfigure}
	\hfill
	\begin{subfigure}[t]{0.45\textwidth}
		\centering
		\includegraphics[width=0.9\linewidth]{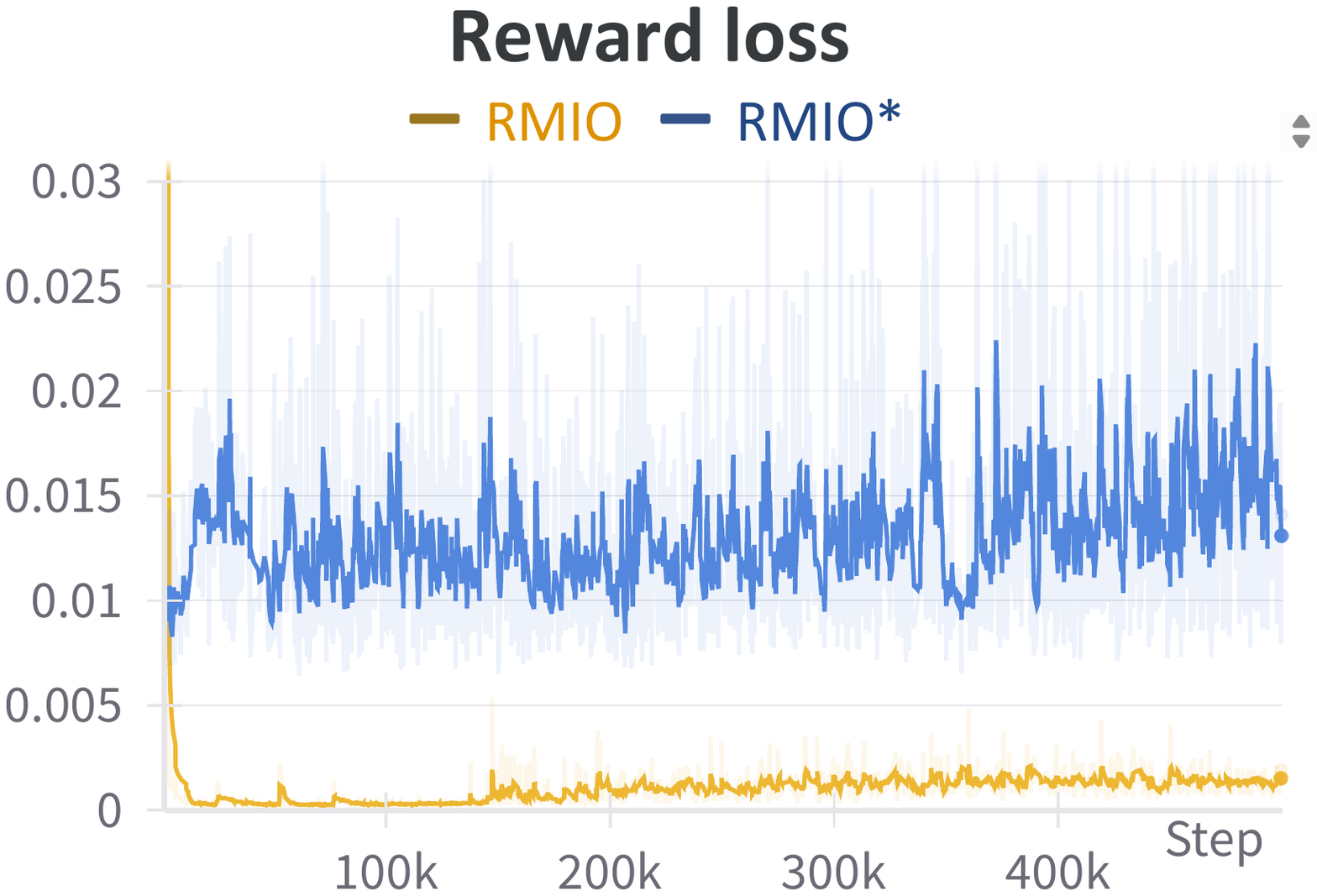}
		\caption{Comparison of reward modeling loss curves between RMIO and RMIO* on the \textit{3s\_vs\_3z} map. The curve is smoothed with EMA and the parameter is set to 0.5.}
		\label{fig:RewardLoss}
	\end{subfigure}
\end{figure}

Ablation experiments on reward smoothing were conducted to demonstrate the superiority of RMIO in reward modeling. The experimental results showed that the loss function of RMIO in reward modeling was significantly smaller than that of its ablation baseline without reward smoothing. For instance, Figure~\ref{fig:Rewardvalue} shows the reward function values of RMIO (with reward smoothing) and RMIO* (without reward smoothing) on the \textit{3s\_vs\_3z} map, while Figure~\ref{fig:RewardLoss} presents the corresponding loss curves. The results indicate that the loss curve with EMA reward smoothing is approximately one-tenth of that in the ablation baseline, and its distribution is also more stable. This demonstrates that reward smoothing can effectively enhance the performance of reward modeling. A detailed discussion of the asymptotic performance of RMIO with reward smoothing will be provided in Section~\ref{sec:asymptotic}.
\subsection{Experiments of Convergence Performance in Standard Environment}
\label{sec:asymptotic}
\begin{figure*}[htbp]
	\centering
	\includegraphics[width=\linewidth]{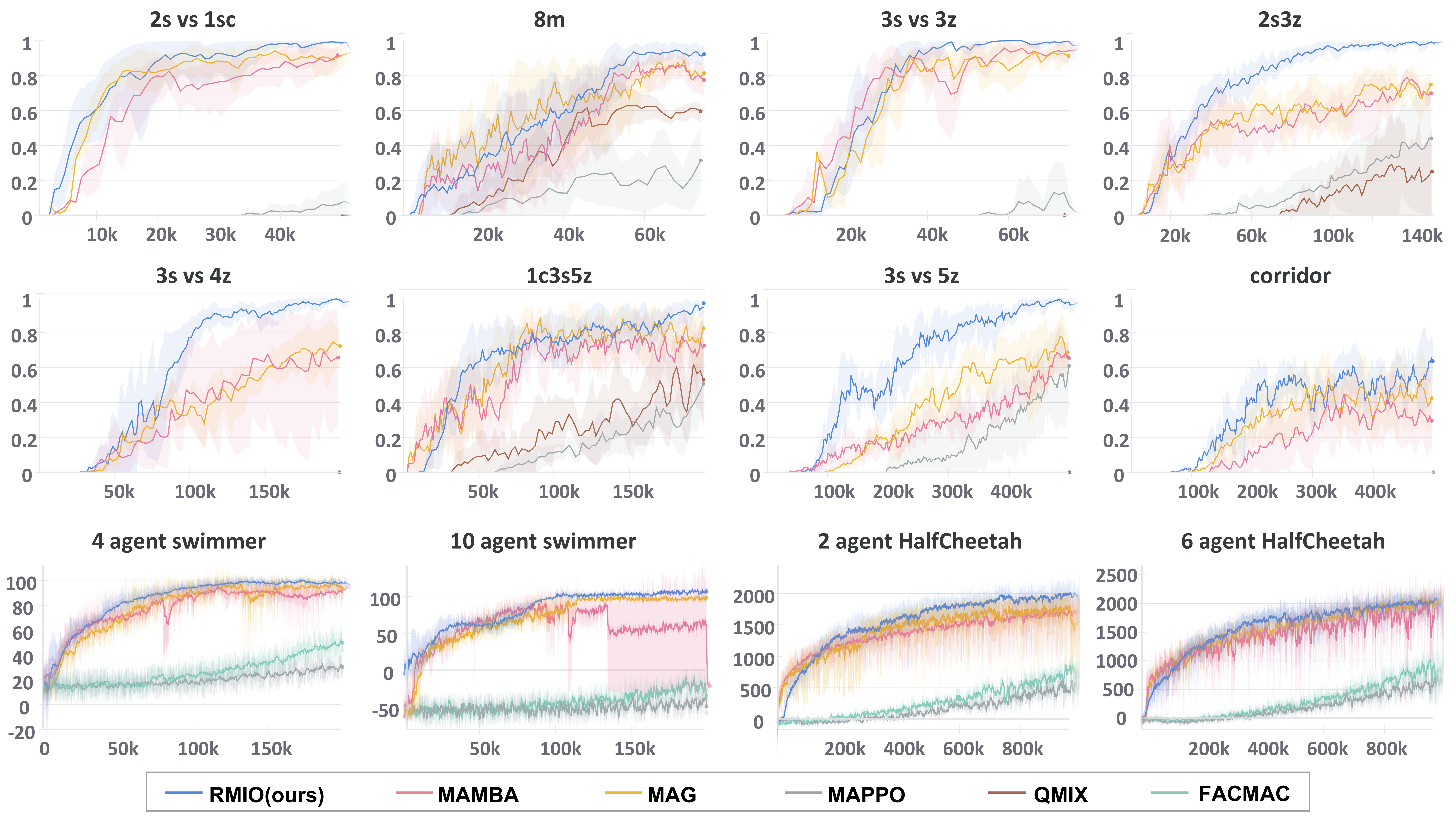}
	\caption{Comparisons with other baselines. The solid line represents the running average of 3 different random seeds, and the shaded area corresponds to the standard deviation between these runs. The X-axis represents the number of steps taken in the real environment, and the Y-axis represents the win rate (SMAC) or episode reward (MAMuJoCo).}
	\label{fig:result}
\end{figure*}
	\vspace{2em}
\begin{table*}[]
\centering
\setlength{\tabcolsep}{1mm}{%
	\begin{tabular}{c|c|c|ccccc}
		\hline
		\multicolumn{8}{c}{\textbf{SMAC}}  \\  \hline
		Maps		    & MES  	 & REIS  & RMIO            
		
		& MAG    & MAMBA    & MAPPO	   & QMIX \\ \hline
		2s\_vs\_1sc    &\multirow{8}{*}{300}	 & 50k		  & \textbf{94(5)}   & 90(8)  & 91(15)   & 9(12)    & 0(0)    \\
		8m             &						 & 75k        & \textbf{94(3)}   & 87(9)  & 77(10)   & 32(15)   & 63(9)    \\
		3s\_vs\_3z     &						 & 75k        & \textbf{96(3)}   & 93(6)  & 90(10)   & 12(16)   & 0(0)    \\
		2s3z  	       &                         & 150k       & \textbf{98(1)}   & 81(11) & 78(13)   & 46(23)   & 23(17)    \\
		3s\_vs\_4z     &   			             & 200k       & \textbf{96(1)}   & 71(13) & 63(37)   & 0(0)     & 0(0)    \\
		1c3s5z         & 						 & 200k       & \textbf{95(3)}   & 78(8)  & 70(9)    & 52(17)   & 46(19)  \\    
		3s\_vs\_5z     &						 & 500k       & \textbf{95(2)}   & 65(15) & 65(8)    & 58(21)   & 0(0)    \\
		corridor       & 						 & 500k       & \textbf{66(19)}  & 55(25) & 46(16)   & 0(0)     & 0(0)    \\ \hline
		\multicolumn{8}{c}{\textbf{MaMuJoCo}}  \\  \hline
		Scenarios		       & MES          & REIS  		     & RMIO                & MAG       & MAMBA    & MAPPO   & FACMAC \\ \hline
		4 agent Swimmer  	   &\multirow{4}{*}{300}    & 200k                   & \textbf{98(3)}      & 94(4)     & 91(6)    & 52(7)   & 32(6)    \\
		10 agent Swimmer       &                        & 200k         		     & \textbf{102(2)}     & 96(3)     & 67(29)   & -20(7)  & -38(6)    \\ 
		2 agent HalfCheetah    &                        & 1m                     & \textbf{2042(48)}   & 1814(58)  & 1759(64) & 823(98) & 531(102)    \\
		6 agent HalfCheetah    & 						& 1m  				     & \textbf{2098(42)}   & 1957(47)  & 1836(58) & 946(104)& 620(94)    \\ \hline
		
	\end{tabular}%
}
\caption{During the training process, the maximum episode steps (MES) is fixed for each map and scene. 
	After completing training for a specified number of real environment interaction steps (REIS) in different environments, the model weights are saved, and the average win rate (in SMAC) or episode reward (in MaMuJoCo), along with their standard deviations, are independently evaluated over 1000 test episodes. Bold numbers highlight the highest average performance among all CTDE methods. RMIO consistently achieves the best performance across all tests. Due to the fact that QMIX is only applicable to discrete environments such as SMAC, the FACMAC method was selected as the model-free method for comparison in the MaMuJoCo environment.
}
\label{tab:experiments}
\end{table*}	
\subsubsection{Baselines} 
We compare RMIO with model-based and model-free baseline methods to assess the convergence performance of our approach under fully observed conditions. The model-based methods include 1) MAMBA, a multi-agent adaptation of DreamerV2~\cite{Hafner_Lillicrap_Norouzi_Ba_2020}, which improves the sample efficiency of MARL by an order of magnitude for the first time; 2) MAG~\cite{Wu_Yu_Chen_Hao_Zhuo_2023}, which is based on MAMBA and takes into account the long-term joint effects of local predictions at each step to generate trajectories with lower cumulative errors, thereby improving the stability of asymptotic convergence performance. The advanced model-free methods include 1) MAPPO~\cite{yu2022surprising}, 2) QMIX~\cite{rashid2020monotonic} and 3)FACMAC~\cite{peng2021facmac}.

\subsubsection{Results and Analysis}
The comprehensive experimental results illustrate the superiority of our approach over both model-based and model-free methods in all test maps or scenarios within a constrained number of iterations, as shown in Table~\ref{tab:experiments} and Figure~\ref{fig:result}. Compared to other CTCE baselines, RMIO, as a CTDE method, achieves a significantly higher win rate on various maps in SMAC environments and achieves higher episode rewards in MaMuJoCo, all while maintaining stable performance across different random seeds and exhibiting markedly better stability in policy convergence.

This advantage primarily stems from the following mechanisms: during the training of the world model, dynamic fluctuations in the distribution of real data samples (e.g., due to sampling bias or exploration during training) may lead to non-stationarity in the world model's performance. During such periods, the output of the world model may deviate significantly from the training data distribution, entering an unacceptable performance regime. CTCE methods (such as MAMBA and MAG) rely on feature vectors generated by the world model as inputs to the policy model. Consequently, when the world model's performance becomes non-stationary, the performance of policy model which is strongly coupled with the world model is more susceptible to degradation, such as the training result of MAMBA in 10 agent swimmer in Figure~\ref{fig:result}. In contrast, RMIO introduces mechanisms such as reward smoothing and a dual-layer experience replay buffer to effectively mitigate the impact of the world model's non-stationary fluctuations on policy training. Furthermore, RMIO reconstructs observation data to serve as inputs to the policy model, a design that reduces the entropy of the input features and thus minimizes the adverse effects of the world model's performance fluctuations on the policy model. Experimental results indicate that, owing to these designs, RMIO significantly outperforms other methods in terms of stability in the MaMuJoCo environment.

\subsection{Experiments of Performance Preservation in Observation-loss Environment}
\subsubsection{Baselines}The lack of world models to fill in missing observation renders model-free methods ineffective in scenarios where observations are incomplete. Although there are some attempts to use traditional prediction methods like Kalman filters and Gaussian predictions in multi-agent environments~\cite{WANG2021106628,LI2023108326}, they struggle to handle the highly non-linear, decentralized, and partially observable complexities, along with the high interaction dynamics, typical of scenarios like SMAC. Hence, only model-based MARL methods are considered for comparison. In the context of ablation experiments, we denote the RMIO method that excludes correction blocks as $\text{RMIO}^*$ for comparative analysis. To ensure fairness in the comparison with MAMBA and MAG, their prior models and observation predictors are employed for predicting and filling in missing observations, aligning with $\text{RMIO} ^ * $.
\subsubsection{Observation-loss Environment Setting}
To assess the performance stability of RMIO in environments affected by observation loss, we introduce an observation loss mechanism based on SMAC and MaMuJoCo. At each time step, there exists a probability \( p_{loss} \) that results in a random subset of agents losing observation information. Notably, to enhance the experimental complexity, we extend the duration of observation loss, that means after the initial occurrence of observation loss, subsequent losses will persist for the following several steps. In the experiment, this duration was set to 10 steps. This setting is also closer to the realistic environmental conditions where the observation environment is harsh (there are strong interference and other factors lasting for a certain period of time). What's more, to intensify the challenge posed by observation loss in easy maps, the number of agents experiencing observation loss in the four maps of \textit{2s\_vs\_1sc, 2s3z,3s\_vs\_3z,3s\_vs\_4z} is set to a fixed number $n-1$ instead of a random number.
\subsubsection{Results and Analysis}
\begin{figure*}[]
	\centering
		\centering
		\includegraphics[width=\linewidth]{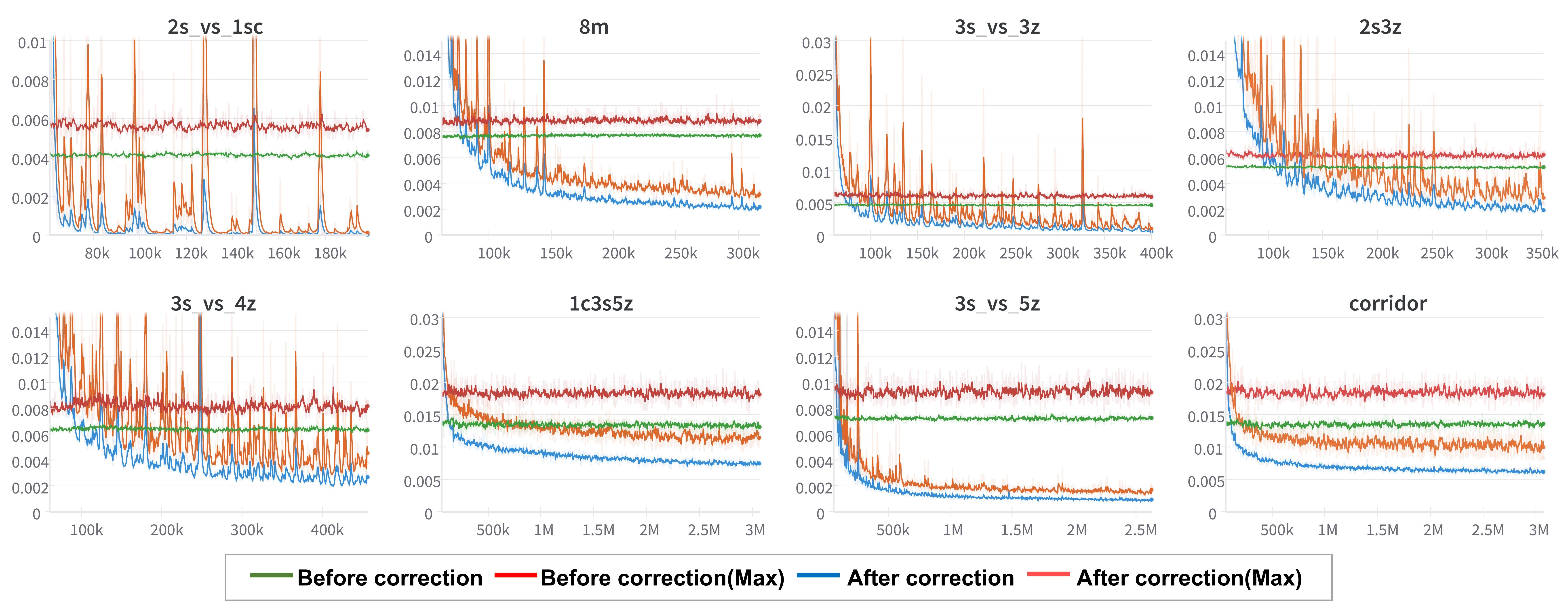}
		\caption{Comparison of loss function values before and after correcting the world model predictions. Before/after correction (MAX) is the maximum value of the correction loss.}
		\label{fig:traincorrection}
	\vspace{2em}
		\centering
		\includegraphics[width=0.99\linewidth]{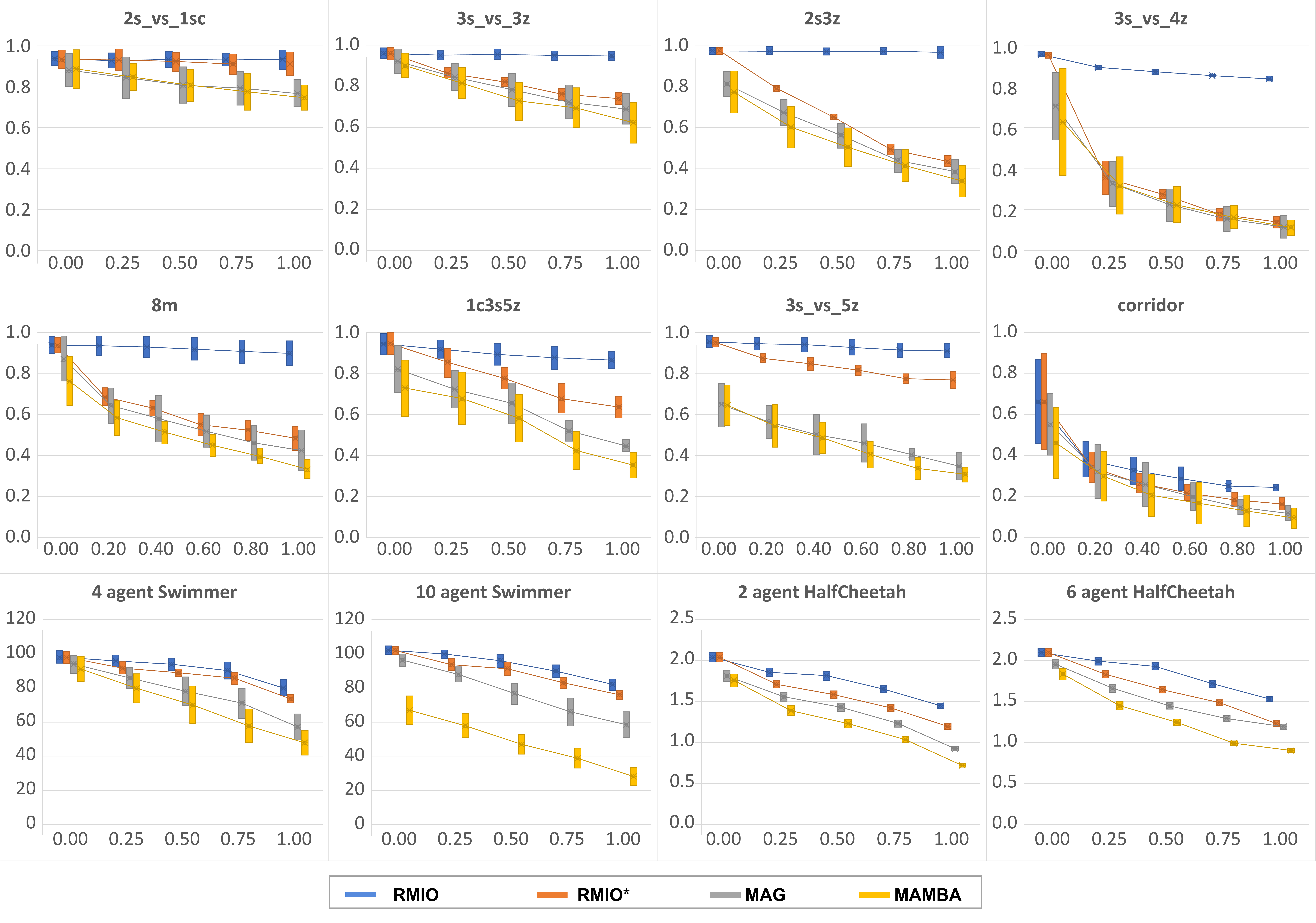}
		\caption{The performance of RMIO method and its ablation control group, as well as other model-based methods, in 1000 random matches under different observation loss probabilities \(p_{loss} \) on different maps in SMAC and different scenarios in MAMUJOCO. In SMAC, the statistical measure is the average percentage win rate and standard deviation of 3 random seeds, while in MaMuJoCo, the statistical measure is the average episode reward and standard deviation of 3 random seeds.}
		\label{fig:correctionresult}

\end{figure*}
Within this environment, the correction block can extract pertinent information from inter-agent communication during training, thereby significantly decreasing observation loss post-calibration. For instance, the training loss curve in the \textit{3s\_vs\_3z} of SMAC is provided in Figure~\ref{fig:traincorrection}. Upon completion of training with 400k real-time steps (equivalent to 2.5M generated time steps), the MSE loss value before correction is 0.007212 and reduced to 0.0009924 after correction. This result shows that the correction block can effectively reduce the prediction error and achieve nearly an order of magnitude reduction. 

After completing the training process, RMIO is tested in the observation-loss SMAC and MaMuJoCo environment. All experimental outcomes are presented in Figure~\ref{fig:correctionresult},  Table~\ref{tab:littleincomplete} and Table~\ref{tab:littleincomplete2} as illustrative examples. The experiment results indicate that the RMIO approach consistently achieves both notably higher win rates across all maps in SMAC and notably higher episode rewards in all scenarios in MaMuJoCo compared to other methods, including the RMIO* ablation comparison method, under varying observation loss probabilities \( p_{loss} \). Moreover, RMIO demonstrates robust performance as the observation loss probability \( p_{loss} \) increases, showing minimal susceptibility to observation loss effects across diverse maps. Remarkably, even when \( p_{loss} = 1 \), RMIO maintains a relatively high win rate in SMAC, particularly in moderately challenging maps. These results suggest that the completion and correction block can effectively estimate missing observations by utilizing temporal recursive relationships and integrating agent-specific information, even in scenarios where observation loss occurs at each time step.

However, it is also observed that while RMIO demonstrates high efficiency in low-difficulty tasks, its efficacy diminishes in high-difficulty tasks (e.g., \textit{corridor}). This decline in performance can be attributed to various factors. Firstly, as the number of agents increases, each agent faces difficulties in observing the states of all other agents, thus weakening the effectiveness of the correction block's fusion of agent-related information. Secondly, in high-difficulty tasks, the accuracy of the prior model's predictions also decreases. This decline in accuracy may stem from the increased complexity of predictions due to higher feature dimensions and the heightened complexity of the joint policy space resulting from a larger number of opponent agents.
\section{Conclusion}
In this work, RMIO first effectively addresses the challenge of lost observation in dynamic environments by leveraging prior prediction models and observation reconstruction predictors to manage missing observation data.  Based on this, a correction block further refines the observation estimates by incorporating correlation information among agents. Moreover,  By decoupling the world model and the policy model, RMIO achieves the CTDE paradigm in standard environment settings. In the case of observation loss, it also only requires limited one-step communication to assist decision-making, while ensuring that the communication frequency is lower than the frequency of observation loss. Additionally, RMIO enhances asymptotic convergence performance through reward smoothing, double replay buffer structure design, and the integration of an additional RNN network in the policy model. Empirically, we show that RMIO outperforms both model-based and model-free baselines on several challenging tasks in the SMAC and MaMuJoCo benchmarks, especially when faced with incomplete observations. Future research will focus on enhancing its robustness by tackling non-stationary performance challenges caused by observation loss in more complex and dynamic scenarios. Additionally, plans are underway to extend its applicability to collaborative competition tasks, with the ultimate goal of developing a comprehensive and adaptive MARL framework that can address a wide range of real-world applications with varying levels of observation completeness and environmental complexity.

\bibliographystyle{elsarticle-num} 
\bibliography{references.bib}


\clearpage

\end{document}